\documentclass[letterpaper]{article} 
\usepackage{aaai23}  
\usepackage{times}  
\usepackage{helvet}  
\usepackage{courier}  
\usepackage[hyphens]{url}  
\usepackage{graphicx} 
\usepackage{natbib}  
\usepackage{caption} 
\usepackage{algorithm}
\usepackage{algorithmic}

\usepackage{newfloat}
\usepackage{listings}
\DeclareCaptionStyle{ruled}{labelfont=normalfont,labelsep=colon,strut=off} 
\floatstyle{ruled}
\newfloat{listing}{tb}{lst}{}
\floatname{listing}{Listing}
\title{Towards a safe MLOps Process for the Continuous Development and Safety Assurance of ML-based Systems in the Railway Domain}
\author{
    Marc Zeller, \textsuperscript{\rm 1}
    Thomas Waschulzik, \textsuperscript{\rm 2}
    Reiner Schmid\textsuperscript{\rm 1},
    Claus Bahlmann, \textsuperscript{\rm 3}
}
\affiliations{
    \textsuperscript{\rm 1} Siemens AG, Otto-Hahn-Ring 6, 81739 Munich, Germany\\
    \textsuperscript{\rm 2} Siemens Mobility GmbH, Siemenspromenade 4, 91058 Erlangen, Germany\\
    \textsuperscript{\rm 3} Siemens Mobility GmbH, 	Rudower Chaussee 29, 12489 Berlin, Germany\\    

    \{forename.surname\}@siemens.com
}

\usepackage{bibentry}

\begin{document}

\maketitle

\begin{abstract}
Traditional automation technologies alone are not sufficient to enable driverless operation of trains (called Grade of Automation (GoA) 4) on non-restricted infrastructure. The required perception tasks are nowadays realized using Machine Learning (ML) and thus need to be developed and deployed reliably and efficiently.
One important aspect to achieve this is to use an MLOps process for tackling improved reproducibility, traceability, collaboration, and continuous adaptation of a driverless operation to changing conditions.
MLOps mixes ML application development and operation (Ops) and enables high frequency software releases and continuous innovation based on the feedback from operations. In this paper, we outline a safe MLOps process for the continuous development and safety assurance of ML-based systems in the railway domain. It integrates system engineering, safety assurance, and the ML life-cycle in a comprehensive workflow. We present the individual stages of the process and their interactions. Moreover, we describe relevant challenges to automate the different stages of the safe MLOps process.
\end{abstract}

\section{Introduction}
With the introduction of driverless train operation, the so-called \emph{Grade of Automation (GoA) 4} on non-restricted infrastructure, the attractiveness of railway systems can be increased significantly. This includes the densification of the timetable, e.g.,~by splitting vehicles that would otherwise run in multiple traction or increased flexibility in timetable design to achieve demand-oriented train services.
Traditional automation technologies alone are not sufficient to enable the driverless operation of trains. However, \emph{Artificial Intelligence (AI)} and \emph{Machine Learning (ML)} offer great potential to realize the mandatory novel functions to replace the tasks of a human train driver, such as obstacle detection on the tracks using ML techniques for computer vision \cite{lecun2015deep}.
%
Assuring safety for a driverless system in complex scenarios on non-restricted railway infrastructure is still unsolved in general. Further, we see the dilemma that the most promising path to a solution is currently seen in using ML approaches, however, ML is imposing additional challenges to formally assuring safety. In the following, we assume that ML components are part of the approach.

Under this assumption, we foresee that different tasks need to be addressed, such that driverless systems for rail can be approved for operation, including
\begin{enumerate}
    \item Systems view: Addressing systems aspects, including linking formal requirements originating from functional safety (e.g., originating from EN 50126\cite{50126}), to the Automated Driving System (ADS) and formalizing a sound safety case 
    \item Safe AI principles and tools: Providing insight into the ML behavior and how it relates to data and further to the requirements
    \item Safe MLOps cycle: Addressing the challenge that ADS also in rail will operate in an open world, which is difficult to specify a-priori and prone to changes during its lifecycle, hence requires agile MLOps cycles including testing \& validation in the field
\end{enumerate}
The safe.trAIn\footnote{https://safetrain-projekt.de/en} project aims to lay the foundation for the safe use of AI/ML to achieve the driverless operation of rail vehicles. 
The project goals are to develop guidelines and methods for the quality assured development of ML components that are able to support the safety assurance of ML in driverless train operation. Based on the requirements for the homologation process in the railway domain, safe.trAIn creates a safety argumentation for an AI-based obstacle detection function of a driverless regional train. Therefore, the project investigates methods to develop  trustworthiness AI-based functions taking data quality, robustness, uncertainty, and transparency/explainability, aspects of the ML model into account. Moreover, the methods need to be integrated into a comprehensive and continuous development, verification, and assessment process for driverless trains.
In this contribution, we focus on the third aspect mentioned earliner. We have other workstreams in the safe.trAIn project along aspects 1 \& 2, for the sake of focus we will address them not in this paper.

While there has been some work on the safety assurance of different ML techniques including Deep Neural Networks (DNNs), it is often neglected, that the system incorporating the ML model must be developed and deployed reliably, efficiently, and in a transparent and traceable way. Especially, since an ML model in a safety-related context needs a rigorous development process like any kind of safety-relevant software. However, there are currently no safety standards available, which describe a development or assurance process for safety-critical ML-based systems.

Traditionally, safety-critical system incorporating software in the railway domain are developed, tested, and validated during design. When the software is updated after the product release, this is done only in a time interval of several months or years, since the re-assessment process takes a lot of time. Therefore, the safety standards EN 50126-1 \cite{50126}, EN 50128 \cite{50128}, EN 50657 \cite{50657}, and EN 50129 \cite{50129} provide both a development process and guidelines for safety assurance. 
In contrast to traditional software, which ensures the correctness in every corner case, ML models are created based on a data set, which can not cover all possible corner cases.
Thus, it is very likely that the ML model will be updated continuously over the product lifetime. There are multiple reasons for this \cite{9320421}: (1) changes necessary due to detected anomalies during operation; (2) changes due to aging of demand of the ML-based system (e.g.~change of the operating context/environment such as domain drift); (3) change in legislation (e.g.~new test cases get mandatory, which leads to an update (re-training) of ML model).
Therefore, an \emph{MLOps} process is required which allows the continues development, verification, and safety assessment of the ML-based functions of a driverless train.

MLOps is a close relative to DevOps. DevOps mixes the development (Dev) and operation (Ops) phases of a software product by promoting high frequency software releases which enable continuous innovation based on feedback from operations \cite{loukides2012devops}. MLOps is an ML engineering culture and practice that aims at unifying ML application development (Dev) and ML application operation (Ops). 

In this paper, we outline a so-called \emph{safe MLOps} process for the continuous development and safety assurance of ML-based systems in the railway domain. It integrates system engineering, safety assurance, and ML life-cycle into a comprehensive and continuous process. Moreover, we describe relevant challenges to speed-up the development of ML-based systems by automating each stage of the safe MLOps process. 

The rest of the paper is organized as follows: In the next section, we briefly summarize relevant related work which provides a foundation for our safe MLOps process. Afterwards, we introduce the safe MLOps process for ML-based systems in the railway domain as defined in the safe.trAIn project. 
At the end, we summarize the main results of the paper and provide an outlook on future research work.

\section{Related Work}
Promoted by the internet companies, MLOps\footnote{https://cloud.google.com/architecture/mlops-continuous-delivery-and-automation-pipelines-in-machine-learning} establishes a DevOps methodology for machine learning applications and is today the de-facto standard for unifying ML model development, deployment, and operation. However, existing MLOps approaches do not address system-level aspects of ML models \cite{NIPS2015_86df7dcf} as well as safety assurance related activities. 

Although there is no standardized process for the development and assurance of ML-based systems in the railway domain, there are already approaches in other domains.

The \emph{UL 4600} (\emph{Evaluation of Autonomous Products}) standard \cite{ul4600} focuses on autonomous driving. It covers the safety principles, tools, and techniques, to design and develop fully automated products. The UL 4600 deﬁnes topics that must be addressed to create a safety case for an autonomous product, but it does not recommend speciﬁc technologies or methods to be used. Moreover, it is based on the development and assurance lifecycle of the ISO 26262-2 standard \cite{iso26262} and does not define explicit ML development or assurance stages.

The whitepaper "Safety First for Autonomous Driving" \cite{wood2019safety} produced by 11 automotive companies and key technology providers illustrates a development and validation process for Deep Neural Networks (DNNs) consisting of the four stages "Define", "Specify", "Develop and Evaluate", and "Deploy and Monitor". In each of the stages, safety artifacts are created which support the safety case. In the whitepaper the challenges that need to be addressed in each of the activities are listed. However, the paper only addresses the development and assurance of an ML model and does not put this into the context of engineering the system. On the other hand, \cite{9320421} presents a detailed process for the development and assurance of ML models, which can be incorporated in the development and assurance process specified in the ISO 26262-2 standard for automotive systems. But the approach focuses on a V-based development process and does not take aspects of MLOps into account to enable continuous or agile product development.

In the avionic domain, the EASA concept paper \cite{EASA} outlines a W-shaped learning assurance process. 
The aim of those systematic actions is to substantiate that errors in a data-driven learning process have been identified and corrected such that the system satisfies the applicable requirements at an adequate level of confidence. However, the paper does not provide any details how to integrate the data-driven learning process into the system development and safety assessment process in the avionic domain (ARP 4754A \cite{arp4754} \& ARP 4761 \cite{arp4761}).

With CRISP-ML(Q) \cite{make3020020} there is a proposal for an iterative process model for ML models which includes a quality assurance methodology. However, this approach was not designed for safety-relevant systems and solely focuses on quality assurance methods to mitigate risks in the development of the ML model.

A process model for machine learning in the context of the Software development process has been captured as a formal process model in \cite{ritz2022}. It considers the interaction of ML model development with the software development process and the required continuous evolution after initial training and testing.
Though safety assurance has not been taken into account, this provides a starting point for defining MLOps in the context of a system development, and the definition of automated pipeline concepts based on the identified interaction points.

Furthermore, there is a domain-independent approach for a development and assurance process of safety-relevant and ML-based autonomous systems by the University of York. The so-called \emph{Assurance of Machine Learning for use in Autonomous Systems (AMLAS) Process} consists of a 6-stage ML lifecycle with the stages "ML safety assurance scoping", "ML safety requirements elicitation", "data management", "model training", "model verification", and "deployment". For each activity safety assurance requirements are listed and potential methods that support each requirement are discussed. Moreover, AMLAS establishes the fundamental link between system-level hazard and risk analysis and component-level safety requirements. Since AMLAS solely focuses on the safety assurance of the ML model, the integration into the system engineering process is not covered.

In \cite{wozniak2021ai} a process model for the engineering of DNNs with the scope of supporting trustworthiness assurance is presented. This process extends the VDE AR E 2842-61-2 standard \cite{VDE}  which outlines a generic framework for the development of trustworthy autonomous systems. The paper describes different development steps for DNNs and illustrates potential methods for trustworthiness assurance in each of the stages. It also outlines how the development and assurance of DNNs links to the development and assurance of the overall system. This approach follows the standard V-model and is not compatible with iterative or agile design practices.

In \cite{waschulzik99} the QUEEN development method is defined for quality assured development of feed forward neural networks. This process combines system engineering approaches to develop appropriate prepossessing stages to reduce the complexity of the task that has to be solved by the machine learning component. To assure that the prepossessing is really reducing the complexity of the given task, complexity measures are introduced for supervised learning data sets. This complexity measures can only be applied if the number of dimensions of the input space is beyond about 10.000 dimensions and if the inputs are at least somehow related to the desired output. The measures give additional support for architectures using the combination of conventional prepossessing with machine learning. The reduction of the complexity of the problem that has to be solved by the ML component reduces the challenge to explain the function that is implemented by the ML model. Additionally, QUEEN introduces local quality indicators that support the quality assurance of data sets. These quality indicators may also be integrated into the MLOps process to support the quality assurance of supervised learning data sets after preprocessing. In \cite{860756} an example is given, how quality assurance problems may be detected in data sets using QUEEN quality indicators.

Based on existing building blocks for ML life-cycle and safety assurance of ML models, we specified an MLOps framework for ML-based systems in the railway domain, which enables continues development, verification, and safety assessment of driverless trains.  

\section{Safe MLOps Process}
%
In this Section, we outline the safe MLOps process specified in the safe.trAIn project, which integrates system engineering of a train with the ML lifecycle and safety assurance activities to continuously develop and assess a driverless train in terms of safety.

\begin{figure*}[t]
\centering
\includegraphics[width=0.97\textwidth]{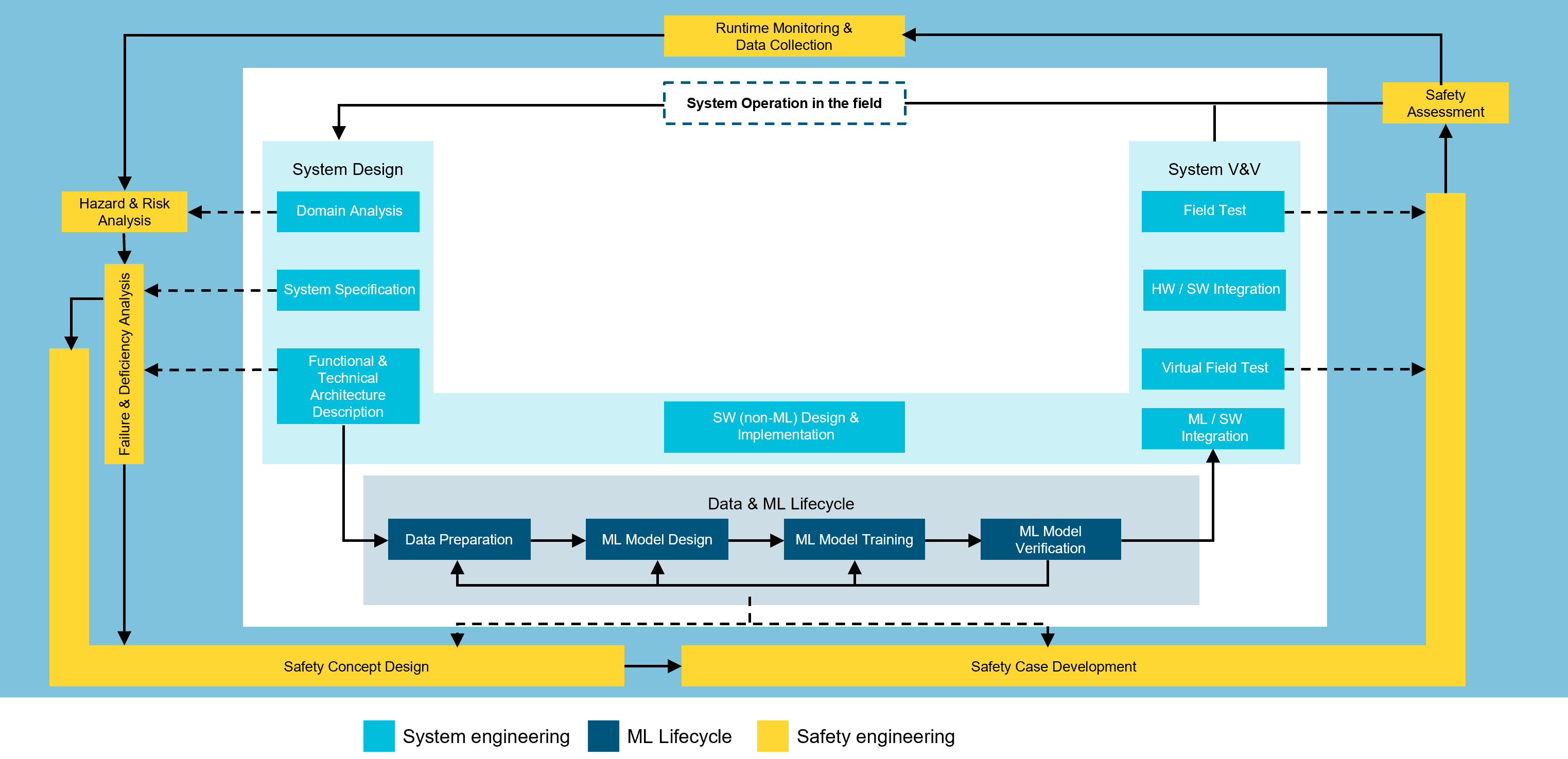}
\caption{Safe MLOps Process for ML-based Systems in the Railway Domain}
\label{fig:process}
\end{figure*}

The safe MLOps process allows continuous engineering and assurance of ML-based systems. Therefore, the individual stages of the process need to be automated and a continuous delivery pipeline from the development of the ML-based system and its components (Dev) to the safety assessment (safe) and afterwards to the operation of the ML-based system (Ops) must be built.

In the "Dev" phase, the engineering of a train is extended with a process to develop and verify ML models which implements functions of the driverless train, such as the obstacle detection. Thereby, existing MLOps approaches are augmented with additional process steps to asses the data quality and assure the quality and performance of the ML model in terms of safety, reliability, transparency, and robustness. Thereby, evidences (quantitative and qualitative) for the safety case are created in each of the stages of the data \& ML lifecycle and incorporated into the overall safety argumentation of the train.

In order to create the safety case in an iterative manner as part of the safe MLOps process, a model-based assurance case based on techniques such as the \emph{Goal Structuring Notation (GSN)} \cite{kelly2004goal} is used to represent the safety argumentation and also the evidences created during the entire development and assurance process.
Our safe MLOps approach aims to adapt the ML model frequently and to re-assess the system which incorporates this ML model in a timely manner. Hence, the ML-based can be adapted continuously to changing operating conditions in an iterative way.

For the verification \& validation of the ML-based system, the ML model is integrated into the overall system and validated using operational test scenarios. In our approach, the ML-based system is tested in different environments. At first, the performance of the individual ML model is tested in isolation (similar to unit tests for non-ML software) using a set of test data during the ML lifecycle in the "Model Verification" stage.
After the integration of ML models and non-ML software, the resulting system is tested in software-in-the-loop (SiL) and hardware-in-the-loop (HiL) environments. The SiL tests are performed in the so-called "Virtual Test Field" stage in our safe MLOps process. After the tests in SiL and HiL environments, ML-based functions of the train are validated during field operation.
After a successful independent "Safety Assessment" based on the safety case with all evidences collected during verification \& validation, the ML-based system (train) is released into operation (the "Ops" phase).

In order to close the loop in the safe MLOps process, relevant parameters of the system are monitored during operation. Thus, new operational conditions must be discovered (e.g., out-of-distribution detection or data/concept drift discovery) and respective data must be collected during the operation of the train. Moreover, in case the system discovers novel situations (not taken into account during the development), the system must handle such situations (e.g., by switching into a safe state) and provide adequate feedback to the next iteration of the development and assurance process with the aim to continuously release an updated and improved version of the ML model.

As depicted in Fig. \ref{fig:process}, the process integrates three parts: (1) the system engineering lifecycle (aligned with the development process of EN 50126-1), (2) the data \& ML lifecycle (aligned with with previous work as outlined in Section 2), and (3) the safety assurance lifecycle (also aligned with the EN 50126-1).

\subsection{System Engineering}
The system engineering process is based on the system lifecycle defined by the EN 50126-1 standard and Model-based System Engineering techniques. It is extended for the engineering of an ML-based system and integrated with the safety engineering lifecycle. The process consists of the following stages:
\begin{enumerate}
    \item Domain Analysis: Apart from the "normal" use case description and requirement elicitation process, also the \emph{Operational Design Domain (ODD)} of the ML-based system is specified. The results of this stage are used as input for the "Hazard \& Risk Analysis (HARA)" stage, which is part of the safety engineering lifecycle.\\
    Challenge: How to specify the ODD in a (semi-formal) way to automatically derive test scenarios for the verification \& validation?
    \item System Specification: Aim of this stage is the specification of the system architecture of the driverless train in form of s SysML model. Based on the system architecture a "Failure \& Deficiency Analysis" is conducted.
    \item Functional \& Technical Architecture Description: In this stage the system architecture is refined into a detailed SW/HW architecture model (in form of a SysML model). The refined architecture is the basis for a refinement of the "Failure \& Deficiency Analysis".
    \item SW (non-ML) Design \& Implementation: For the design and implementation of non-ML SW, the well established practices of the EN 50657 as well as DevOps principles are applied.
    \item ML/SW Integration: In this stage, the non-ML SW and the ML model(s) are integrated to create the \emph{System-Under-Test (SUT)} which is used in the following verification \& validation stages.\\
    Challenge: How to extend existing continuous integration concepts for SW to incorporate also ML models? 
    \item Virtual Test Field: Perform automated integration/system tests for the ML-based system in a SiL environment using the previously built SUT. The evaluation results provide evidences to the "Safety Case Development".\\
    Challenge: Automate Software-in-the-Loop tests, incl. Scenario-based testing, data-driven ODD coverage, and evaluate safety metrics in a simulation environment with both real and synthetic test data.
    \item HW/SW Integration: Deployment of the non-ML SW and the ML model(s) on the target HW and incorporation of physical sensors to perform automated integration/system tests in a Hardware-in-the-Loop environment. The results of these tests provide evidences to the "Safety Case Development".\\
    Challenge: Automation of Hardware-in-the-Loop tests, similar to Software-in-the-Loop tests in the virtual test field.
    \item Field Test: Finally, the ML-based system (train) is validated in specific test scenarios during operation on real test tracks. Again, the validation results provide evidences to the "Safety Case Development".\\
    Challenge: Field testing must be performed in two stages: First, in a so-classed \emph{shadow mode} with the aim to collect data for the validation of the system under failure conditions/scenarios. After successful validation in the shadow mode, system validation tests are performed in an \emph{operative mode}, in which also the actuators of the rolling stock are controlled by the new SW or ML model.
    \item System Operation: After successful independent safety assessments, the train is operating in the defined ODD. Feedback from the operation is provided by collection data, see "Runtime Monitoring \& Data Collection" stage.
\end{enumerate}
Please note, the 9 stages of the system engineering process cover the 12 stages defined in the EN 50126-1 standard. Only the stages "Manufacture" and "Decommissioning" are not represented in our safe MLOps process, since these stages do not need to be adjusted for an ML-based system in the railway domain. Moreover, some of the stages of the system engineering lifecycle cannot be automated such as the definition of a system architecture.

\subsection{Data \& ML Lifecycle}
In addition to the design and implementation of (non-ML) SW, there is a dedicated lifecycle for the ML model including the data for training and verification. The ML lifecycle uses the ODD specification, the technical system architecture, and the (safety) requirements as input. The result of the ML lifecycle is a trained and verified ML model which is input to the ML/SW Integration stage of the system engineering lifecycle. The ML lifecycle consists of the following 4 stages:
\begin{enumerate}
    \item Data Preparation: In this stage, the quality of training and test data used for the training and verification of the model is ensured. Therefore, corner cases and adversarial examples must be collected, training data must be selected, and synthetic data need to be generated. All the data used must be checked w.r.t.~labeling inconsistencies. To analyse the quality of the data, quality indicators based on QUEEN \cite{waschulzik99} are applied.\\
    Challenge: How to automatically analyze the ODD in terms of data distribution and diversity?
    \item ML Model Design: The ML model is defined, e.g.~the number of layers of a DNN is specified. The design of the ML model influence the "Safety Concept", since different types of ML techniques require different safety argumentation strategies. 
    \item ML Model Training: In this stage, the ML model is trained using the data from the "Data Preparation" phase.
    \item ML Model Verification: The performance of the ML model is verified in this stage. Therefore, different KPIs must be evaluated, such as uncertainty, robustness, interpretability, etc. These KPIs are used in the "Safety Case Development" as evidences to support the safety argumentation. In case the defined KPIs/requirements of the ML model are not met, you can go back to any of the ML model lifecycle stages and retrain the model with new data or modify the ML model design.\\
    Challenge: What are the right (quantitative) metrics to evaluate the ML model performance and support the safety argumentation as evidences?    
\end{enumerate}

\subsection{Safety Engineering}
Both the system engineering and the ML lifecycle provide input for the safety engineering lifecycle, which consists of the following 6 stages:
\begin{enumerate}
    \item Hazard \& Risk Analysis (HARA): The HARA tries to identify potential hazards that can be caused by the ML-based system. This stage, equals the \emph{Risk analysis and evaluation} phase of the EN 50126-1 standard. As a result of this step, safety goals are defined as top-level safety requirements, which are an extension of the system requirements.\\
    Challenge: Is it possible to automate both the identification of hazards and the assessment of the risk associated to hazards?
    \item Failure \& Deficiency Analysis: In order to refine the safety requirements and create a safety architecture for a driverless train, the available system specification (system model) is used as input to safety analyses at different abstraction levels in order to identify potential causes of the identified hazards. This is done in parallel with the respective system engineering activities. Since we deal with an ML-based system, we need to incorporate \emph{Safety Of The Intended Functionality (SOTIF)} aspects according to ISO 21448 \cite{iso21448} in the safety analyses. Component Fault \& Deficiency Trees (CFDTs) \cite{cfdt} allow the analysis of both functional safety and SOTIF aspects of the system. Moreover, CFDTs is a Model-based Safety Assurance (MBSA) methodology, which seamlessly integrates with the MBSE approach used for system engineering. Hence, consistency between the system model (in SysML) and the safety analyses (in form of CFDTs) can be ensured despite the iterative development approach.\\
    Challenge: Automate the safety analysis as much as possible using MBSA. Although a full automation without safety experts in the loop is hardly possible.
    \item Safety Concept Design: The safety concept is deﬁned as the speciﬁcation of the safety requirements, their allocation to system elements, and their interactions necessary to achieve safety goals. Based on the complete and detailed system design (including the design of the ML model) and the outcome of the "Failure \& Deficiency Analysis", a concept is created which represents the argumentation that the ML-based train is sufficiently safe. Therefore, an assurance case in form of a \emph{Goal Structuring Notation (GSN)} \cite{kelly2004goal} tree is specified to represent the safety argumentation. It provides the possibility to integrate evidences from the verification \& validation activities in the argumentation tree.\\
    Challenge: Since there is not yet a standard for safety assessment of ML-based systems in the railway domain, the requirements a safety argumentation must fulfill are not clear. The safe.trAIn project tries to fill this gap.
    \item Safety Case Development: In this stage, a sound safety case is created which is the basis for the independent safety assessment of the system's safety.\\
    Challenge: How to automatically adapt and extend the safety case when the ODD or the available data changes?
    \item Safety Assessment: The safety case of the driverless train is subject to independent reviews and safety assessments.\\
    Challenge: How to deliver necessary assets and documentation for the assessment at the end of each sprint/iteration? And how enable the reuse of previous results in an assessment of a changed systems?
    \item Runtime Monitoring \& Data Collection: During operation the ML-based system must be monitored throughout the entire product lifetime to detect potentially novel situations (which were unknown during the development) or potential unsafe behavior.\\
    Challenge: Which are the relevant safety metrics to be continuously monitored during operation in order to identify potential unsafe behavior or novel scenarios?
\end{enumerate}

\section{Summary and Outlook}
%
In this paper, we outline a process for the continuous development and safety assurance of ML-based systems in the railway domain. This so-called \emph{safe MLOps} process integrates system engineering, safety assurance, and the Machine Learning (ML) life-cycle into a comprehensive and continuous process.
While the system engineering and the safety assurance related stages of the process are aligned with the development process of the EN 50126-1 safety standard for railway systems, the ML \& data lifecycle is aligned with previous work in this field (e.g., the AMLAS process \cite{amlas}). Apart from outlining the safe MLOps process and describing the interaction of the different stages of the systems engineering, the ML lifecycle, and the safety assurance, we describe relevant challenges to automate the stages of the safe MLOps process.

As a next step, the described safe MLOps process is realized in the safe.trAIn project in form of a CI/CD pipeline with the appropriate support by tooling to achieve the required degree of automation. The resulting pipeline will be evaluated in an industrial use case. In order to to establish traceability and audibility a consistency layer is required which stores all the artifacts created in the different stages of the safe MLOps process (e.g.~parameter of the ML model, training/test data, evidences, etc.). 
Moreover, solution strategies for the challenges to automate the different stages of the safe MLOps process must be developed in the safe.trAIn project. One of these challenges is keeping the Verification \& Validation test chain in sync with the demands of the safety assurance case.

We will also discuss our safe MLOps process with the relevant standardization bodies, such as CEN-CENELEC JTC 21, with the goal to specify a standardized process for the development and safety assurance of ML-based systems in the railway domain.
Furthermore, we will investigate the applicability of the safe MLOps framework to different industrial application domains such as industrial automation, healthcare, etc.

\section*{Acknowledgment}
This research has received funding from the \emph{Federal Ministry for Economic Affairs and Climate Action} (BMWK) under grant agreements 19I21039A.

\bibliography{aaai23}

\begin{thebibliography}{24}
\providecommand{\natexlab}[1]{#1}

\bibitem[{{EASA}(2021)}]{EASA}
{EASA}. 2021.
\newblock European Union Aviation Safety Agency (EASA) Concept Paper: First
  usable guidance for Level 1 machine learning applications.
\newblock https://www.easa.europa.eu/en/downloads/134357/en.

\bibitem[{{EN 50126-1:2018-10}(2018)}]{50126}
{EN 50126-1:2018-10}. 2018.
\newblock {Railway Applications -- The Specification and Demonstration of
  Reliability, Availability, Maintainability and Safety (RAMS) - Part 1:
  Generic RAMS Process}.

\bibitem[{{EN 50128:2012-02}(2012)}]{50128}
{EN 50128:2012-02}. 2012.
\newblock {Railway Applications -- Communication, signaling and processing
  systems -- Software for railway control and protection systems}.

\bibitem[{{EN 50129:2019-06}(2019)}]{50129}
{EN 50129:2019-06}. 2019.
\newblock {Railway application -- Communications, signaling and processing
  systems -- Safety related electronic systems for signaling}.

\bibitem[{{EN 50657:2017-11}(2017)}]{50657}
{EN 50657:2017-11}. 2017.
\newblock {Railways Applications - Rolling stock applications - Software on
  Board Rolling Stock}.

\bibitem[{Hawkins et~al.(2021)Hawkins, Paterson, Picardi, Jia, Calinescu, and
  Habli}]{amlas}
Hawkins, R.; Paterson, C.; Picardi, C.; Jia, Y.; Calinescu, R.; and Habli, I.
  2021.
\newblock Guidance on the Assurance of Machine Learning in Autonomous Systems
  (AMLAS).
\newblock https://arxiv.org/abs/2102.01564.

\bibitem[{{ISO 21448:2022-06}(2022)}]{iso21448}
{ISO 21448:2022-06}. 2022.
\newblock {Road vehicles - Safety of the Intended Functionality}.

\bibitem[{{ISO 26262-2:2018-12}(2018)}]{iso26262}
{ISO 26262-2:2018-12}. 2018.
\newblock {Road vehicles -- Functional safety -- Part 2: Management of
  functional safety}.

\bibitem[{Kelly and Weaver(2004)}]{kelly2004goal}
Kelly, T.; and Weaver, R. 2004.
\newblock The goal structuring notation--a safety argument notation.
\newblock In \emph{Proceedings of the dependable systems and networks 2004
  workshop on assurance cases}, 6.

\bibitem[{Koopman et~al.(2019)Koopman, Ferrell, Fratrik, and Wagner}]{ul4600}
Koopman, P.; Ferrell, U.; Fratrik, F.; and Wagner, M. 2019.
\newblock A Safety Standard Approach for Fully Autonomous Vehicles.
\newblock In Romanovsky, A.; Troubitsyna, E.; Gashi, I.; Schoitsch, E.; and
  Bitsch, F., eds., \emph{Computer Safety, Reliability, and Security},
  326--332. Springer International Publishing.

\bibitem[{LeCun, Bengio, and Hinton(2015)}]{lecun2015deep}
LeCun, Y.; Bengio, Y.; and Hinton, G. 2015.
\newblock Deep learning.
\newblock \emph{nature}, 521(7553): 436--444.

\bibitem[{Loukides(2012)}]{loukides2012devops}
Loukides, M. 2012.
\newblock \emph{What is DevOps?}
\newblock O'Reilly Media, Inc.

\bibitem[{Radlak et~al.(2020)Radlak, Szczepankiewicz, Jones, and
  Serwa}]{9320421}
Radlak, K.; Szczepankiewicz, M.; Jones, T.; and Serwa, P. 2020.
\newblock Organization of machine learning based product development as per ISO
  26262 and ISO/PAS 21448.
\newblock In \emph{2020 IEEE 25th Pacific Rim International Symposium on
  Dependable Computing (PRDC)}, 110--119.

\bibitem[{Ritz et~al.(2022)Ritz, Phan, Sedlmeier, Altmann, Wieghardt, Schmid,
  Sauer, Klein, Linnhoff-Popien, and Gabor}]{ritz2022}
Ritz, F.; Phan, T.; Sedlmeier, A.; Altmann, P.; Wieghardt, J.; Schmid, R.;
  Sauer, H.; Klein, C.; Linnhoff-Popien, C.; and Gabor, T. 2022.
\newblock Capturing Dependencies Within Machine Learning via a Formal Process
  Model.
\newblock In \emph{Leveraging Applications of Formal Methods, Verification and
  Validation. Adaptation and Learning. ISoLA2022. Lecture Notes in Computer
  Science, vol 13707. Springer}, 249--265.

\bibitem[{{SAE}(1996)}]{arp4761}
{SAE}. 1996.
\newblock {ARP 4761: Guidelines and Methods for Conducting the Safety
  Assessment Process on Civil Airborne Systems and Equipment}.

\bibitem[{{SAE}(2010)}]{arp4754}
{SAE}. 2010.
\newblock {ARP 4754A: Guidelines for Development of Civil Aircraft and
  Systems}.

\bibitem[{Sculley et~al.(2015)Sculley, Holt, Golovin, Davydov, Phillips, Ebner,
  Chaudhary, Young, Crespo, and Dennison}]{NIPS2015_86df7dcf}
Sculley, D.; Holt, G.; Golovin, D.; Davydov, E.; Phillips, T.; Ebner, D.;
  Chaudhary, V.; Young, M.; Crespo, J.-F.; and Dennison, D. 2015.
\newblock Hidden Technical Debt in Machine Learning Systems.
\newblock In Cortes, C.; Lawrence, N.; Lee, D.; Sugiyama, M.; and Garnett, R.,
  eds., \emph{Advances in Neural Information Processing Systems}, volume~28.
  Curran Associates, Inc.

\bibitem[{Studer et~al.(2021)Studer, Bui, Drescher, Hanuschkin, Winkler,
  Peters, and Müller}]{make3020020}
Studer, S.; Bui, T.~B.; Drescher, C.; Hanuschkin, A.; Winkler, L.; Peters, S.;
  and Müller, K.-R. 2021.
\newblock Towards CRISP-ML(Q): A Machine Learning Process Model with Quality
  Assurance Methodology.
\newblock \emph{Machine Learning and Knowledge Extraction}, 3(2): 392--413.

\bibitem[{{VDE-AR-E 2842-61-2 Anwendungsregel:2021-06}(2021)}]{VDE}
{VDE-AR-E 2842-61-2 Anwendungsregel:2021-06}. 2021.
\newblock Development and trustworthiness of autonomous/cognitive systems.

\bibitem[{Waschulzik(1999)}]{waschulzik99}
Waschulzik, T. 1999.
\newblock \emph{Qualit{\"a}tsgesicherte effiziente Entwicklung
  vorw{\"a}rtsgerichteter k{\"u}nstlicher Neuronaler Netze mit {\"u}berwachtem
  Lernen (QUEEN)}.
\newblock Dissertation, Technische Universit{\"a}t M{\"u}nchen, Fachbereich
  Informatik.

\bibitem[{Waschulzik et~al.(2000)Waschulzik, Brauer, Castedello, and
  Henery}]{860756}
Waschulzik, T.; Brauer, W.; Castedello, T.; and Henery, B. 2000.
\newblock Quality assured efficient engineering of feedforward neural networks
  with supervised learning (QUEEN) evaluated with the"pima indians diabetes
  database".
\newblock In \emph{Proceedings of the IEEE-INNS-ENNS International Joint
  Conference on Neural Networks. IJCNN 2000. Neural Computing: New Challenges
  and Perspectives for the New Millennium}, volume~4, 97--102 vol.4.

\bibitem[{Wood et~al.(2019)Wood, Robbel, Maass, Tebbens, Meijs, Harb, Reach,
  Robinson, Wittmann, Srivastava et~al.}]{wood2019safety}
Wood, M.; Robbel, P.; Maass, M.; Tebbens, R.~D.; Meijs, M.; Harb, M.; Reach,
  J.; Robinson, K.; Wittmann, D.; Srivastava, T.; et~al. 2019.
\newblock Safety first for automated driving.
\newblock
  https://newsroom.intel.com/wp-content/uploads/sites/11/2019/07/Intel-Safety-First-for-Automated-Driving.pdf.

\bibitem[{Wozniak, Putzer, and C{\^a}rlan(2021)}]{wozniak2021ai}
Wozniak, E.; Putzer, H.~J.; and C{\^a}rlan, C. 2021.
\newblock AI-Blueprint for Deep Neural Networks.
\newblock In \emph{SafeAI@ AAAI}.

\bibitem[{Zeller(2022)}]{cfdt}
Zeller, M. 2022.
\newblock Component Fault and Deficiency Tree (CFDT): Combining Functional
  Safety and SOTIF Analysis.
\newblock In \emph{Model-Based Safety and Assessment}, 146--152.
\newblock ISBN 978-3-031-15842-1.

\end{thebibliography}

\end{document}